\documentstyle[12pt]{article}
\begin{document}
\baselineskip 22pt plus 2pt

\begin{flushright}
TECHNION-PH-96-7
\end{flushright}
\vspace{2cm}

\begin{center}
{\bf Asymmetry in the decay $\Omega^-\rightarrow\Xi^-\gamma$}\\

G. Eilam, A. Ioannissyan$^{*}$ and P. Singer\\

Department of Physics,\\ Technion - Israel Institute of Technology, Haifa 
32000, Israel
\end{center}

\noindent{\bf Abstract}

We consider the asymmetry in the decay $\Omega^-\rightarrow\Xi^-\gamma$
assuming that a Vector Meson Dominance approach for the $s\rightarrow d\gamma$ 
transition gives
the dominant contribution.  Since in this long-distance approximation the decay
is due to a single quark transition $s\rightarrow d\gamma$, the angular distribution
asymmetry is given by the single positive asymmetry parameter 
$\alpha_h = \frac{M^2_s-M^2_d}{M^2_s+M^2_d} = 0.4\pm0.1$.  We also discuss the
asymmetry in $\Xi \rightarrow \Sigma^-\gamma$, which is expected to be between
-0.2 and 0.3.
\vfill

\noindent\rule{6cm}{0.020cm}\\
\noindent $^*$ On leave from Yerevan Physics Institute, Alikhanyan Br. 2, Yerevan,\\ 
\hspace*{0.35cm}375036, Armenia.\\
\pagebreak

The process $\Omega^-\rightarrow\Xi^-\gamma$ has a special place [1-4] among hyperon
radiative decays, since the quark composition of the participating hadrons
precludes W-exchange among pairs of valence quarks to induce this decay.  A similar
situation occurs in $\Xi^-\rightarrow\Sigma^-\gamma$ decay.  Accordingly,
these decays have been singled out [2] as possible candidates for the detection of
the single quark $s\rightarrow d\gamma$ magnetic transition.  Using the present
knowledge on the QCD-corrections to the effective nonleptonic Hamiltonian, the short-distance
electroweak penguin contribution to these rates [1,5,6] turns out to be lower
by about two orders of magnitude than the present experimental upper limit
on the $\Omega^-$ decay rate ($\Gamma^{\Omega^-\rightarrow\Xi^-\gamma}_{\exp}<
(3.7\cdot10^{-9}$eV[7]) and lower by about one order of magnitude than
the experimental value for $\Xi^-\rightarrow\Sigma^-\gamma$
($\Gamma^{\Xi^-\rightarrow\Sigma^-\gamma}_{\exp}=
(5.10\pm0.92)\cdot10^{-10}$eV[8]).  A similar result is given by the calculation
of gluonic penguins [9,10].  Turning to long-distance contributions, the
structure of the nonleptonic $\Delta S = 1$ Hamiltonian does not induce pole
contributions in the $\Omega^-\rightarrow\Xi^-\gamma$ decays.  Kogan and Shifman [1]
have calculated the two-particle intermediate ``s-channel'' contributions to
radiative decays of hyperons.  For the decay $\Omega^-\rightarrow\Xi^-\gamma$
they found $\Gamma^{\Omega^-\rightarrow\Xi^-\gamma}_{\rm "s-channel"} 
\simeq 10^{-10}$eV.  Thus, the ``s-channel'' contributions are lower than the 
present experimental limit by a factor of about 40.  For the decay
$\Xi^-\rightarrow\Sigma^-\gamma$ their result is of the same order as the 
present experimental value.  On the other hand, a Vector Meson Dominance (VMD) 
approach to hyperon radiative decays on the hadronic level [11], which uses
$SU(6)_W$-symmetry to determine the parity-violating couplings of vector mesons
to baryons from nonleptonic hyperon decays, finds a branching ratio for 
$\Omega^-\rightarrow\Xi^-\gamma$ which is already slightly higher than the 
experimental limit.

A different approach [12] for calculating the long-distance contribution, using
a VMD approximation for the $s\rightarrow d\gamma$ transition (along the lines
discussed by Deshpande et al. [13] for $b\rightarrow s\gamma$), shows that
this contribution can be of the order of the present experimental limit.

The $s\rightarrow d\gamma$ amplitude in the VMD approximation is [12]
\begin{eqnarray}
A_{\rm VMD} =
e\frac{G_F}{\sqrt{2}} \sin \theta_ca_2(m^2_s)C_{\rm VMD}
\frac{1}{M^2_s-M^2_d} \bar{d}\sigma^{\mu\nu}
[M_sR-M_dL]sF_{\mu\nu} \ ,           
\end{eqnarray}
\begin{eqnarray}
C_{\rm VMD} = \frac{2}{3}\sum_i\frac{g^2_{\psi_i}(0)}{m^2_{\psi_i}} -
\frac{1}{2}\frac{g^2_\rho(0)}{m^2_\rho} - 
\frac{1}{6}\frac{g^2_\omega(0)}{m^2_\omega} \ ,
\end{eqnarray}
where $\sin\theta_c = 0.22, \ a_2(m^2_s)\geq 0.5$ is a phenomenonologically
determined QCD coefficient, and $g_V^2$ are the couplings of vector mesons
to photons.  Due to the nature of the VMD approximation, $M_s$ and
$M_d$ should correspond to ``constituent'' mass parameters.

The rate of the $\Omega^-$ radiative decay induced by the 
$s\rightarrow d\gamma$ transition of Eq.~(1) is [12,14,15]

\begin{eqnarray}
\Gamma^{\Omega^-\rightarrow\Xi^-\gamma} = 
\frac{16\alpha G^2_F}{3} \left(\frac{m_{\Xi^-}}{m_{\Omega^-}}\right)
|\vec{q}|^3 \sin^2 \theta_c a^2_2 C^2_{\rm VMD}
\frac{M^2_s + M^2_d}{(M^2_s - M^2_d)^2} 
\end{eqnarray}
where $\vec{q}$ is the photon momentum in the $\Omega^-$ rest frame.

The large theoretical uncertainty of over 40\% in the value of the sum
$\Sigma_i\frac{g^2_{\psi_i}(0)}{m^2_{\psi_i}}$ which appears in $C_{\rm VMD}$,
would allow the VMD contribution to saturate the experimental bound.  In fact,
the experimental limit on the decay rate $\Omega^-\rightarrow\Xi^-\gamma$
can be used to constrain $|C_{\rm VMD}|<0.01$GeV$^2$ [12].

>From Eq.~(1) it follows that on the quark level the angular distribution has
the form
\begin{eqnarray}
\frac{1}{\Gamma^{s\rightarrow d\gamma}}
\frac{d\Gamma^{s\rightarrow d\gamma}}{d(\cos\theta)} =
\frac{1}{2}(1+\alpha_h\cos\theta)
\end{eqnarray}
where
\begin{eqnarray}
\alpha_h = \frac{M^2_s - M^2_d}{M^2_s + M^2_d} = 0.4\pm0.1    
\end{eqnarray}
Here $\theta$ is the angle between the spin of the $s$ quark and 
the direction of
the three-momentum of the d quark.  This functional form for $\alpha_h$ is valid 
for the short-distance as well [4,16], but in this case one uses current quark
masses to obtain $\alpha_h\simeq 1$.

The angular distribution for the radiative decay $\Omega^-\rightarrow\Xi^-\gamma$,
when we take into account all possible contributions to this decay, is proportional
to $(\alpha_0+\alpha_1\cos\theta + \alpha_2\cos^2\theta + \alpha_3\cos^3\theta)$.
But it turns out [14,15] that the angular distribution of the decay rate
$\Omega^-\rightarrow\Xi^-\gamma$, when it is going through a single quark 
transition $s\rightarrow d\gamma$, is given by the same single asymmetry parameter
$\alpha_h$ of eq.~(4) and has the following form [14,15]
\begin{eqnarray}
\frac{1}{\Gamma^{\Omega^-\rightarrow\Xi^-\gamma}}
\left. \frac{d\Gamma^{\Omega^-\rightarrow\Xi^-\gamma}}{d(\cos\theta)}\right|_{s^\Omega_z=3/2}
=\frac{3}{8}(1+2\alpha_h\cos\theta+\cos^2\theta)     
\end{eqnarray}
\begin{eqnarray}
\frac{1}{\Gamma^{\Omega^-\rightarrow\Xi^-\gamma}}
\left. \frac{d\Gamma^{\Omega^-\rightarrow\Xi^-\gamma}}{d(\cos\theta)}\right|_{s^\Omega_z=1/2}
=\frac{5}{8}(1+\frac{2\alpha_h}{5} \cos\theta-\frac{3}{5}\cos^2\theta)     
\end{eqnarray}
where $\theta$ is the angle between the direction of the outgoing baryon $\Xi^-$
and the $z$ axis, and $s^\Omega_z = 3/2, \ 1/2$ are the projections of the 
spin of $\Omega$ along the $z$ axis.

As it is noted in ref.~[14] the decay rate of $\Xi^-\rightarrow \Sigma^-\gamma$
has the same form for the angular distribution as the quark decay $s\rightarrow d \gamma$
eq.~(4) i.e. $(1+\alpha_h\cos\theta$).  But in the decay $\Xi^-\rightarrow \Sigma^-\gamma$
there are two types of contributions to the asymmetry parameter as well as to the
decay rate.  The long-distance $s\rightarrow d\gamma$ contribution to the 
rate is 
$\Gamma^{\Xi^-\rightarrow \Sigma^-\gamma}_{\rm VMD} < 4.4\cdot10^{-10}$eV [12],
while the two particle ``s-channel'' contribution is $\sim 5\cdot10^{-10}$eV [1,17].
Taking $C_{\rm VMD} = 0.01$  and the particle loop contribution as in Ref.~[17]
we find $-0.2<\alpha_h<0.3$, the range being the result of the 
unknown phase between the two contributions.  By decreasing $C_{\rm VMD}$ to zero, $\alpha_h$ is
kept in the above range but the branching ratio becomes somewhat too large.

To summarize, we predict a positive asymmetry equal to $0.4\pm0.1$ for the 
decay $\Omega^-\rightarrow \Xi^-\gamma$, based on the $s\rightarrow d\gamma$
VMD approach of ref.~[12].  For the decay
$\Xi^-\rightarrow \Sigma^-\gamma$
we predict an asymmetry in the range $-0.2$ and $0.3$.  We urge the measurement
of these asymmetry parameters which will hopefully elucidate the nature of the 
$s\rightarrow d\gamma$ transition and its role in hyperon radiative decays.\\

\noindent{\bf Acknowledgements}

The research of A.I. and P.S. is supported in part by Grant 5421-2-96 from the 
Ministry of Science and the Arts of Israel.  A.I. has been partially supported
by the Lady Davis Trust.  The work of G.E. and P.S. has been supported in part
by the Fund for Promotion of Research at the Technion and of G.E. by GIF
and by the Israel Science Foundation.\\
\pagebreak

\noindent{\bf References}

\begin{enumerate}
\item Ya. I. Kogan and M.A. Shifman, Sov.\ J.\ Nucl.\  Phys.\ {\bf 38}, 628 (1983).
\item L. Bergstr\"{o}m and P. Singer, Phys.\ Lett.\ {\bf B169}, 297 (1986).
\item P. Singer, Essays in honor of M. Roos (M. Chaichian and J. Maalampi, eds.),
1991, p.~143.
\item For a recent review on the experimental and theoretical status of this
field see J.\ Lach and P. Zenczykowski, Int.\ J.\ Modern Phys.\ {\bf A10},
3817 (1995).
\item M.A. Shifman, A.I. Vainshtein and V.I. Zakharov, Phys.\ Rev.\ {\bf D18},
2583 (1978); M. McGuigan and A.I. Sanda, Phys.\ Rev. {\bf D36}, 1413 (1987).
\item S. Bertolini, M. Fabbrichesi and E. Gabrielli, Phys.\ Lett.\ {\bf B327}, 1361 (1994).
\item I.F. Albuquerque et al., Phys.\ Rev.\ {\bf D50}, 18 (1994).
\item Particle Data Group, Phys. Rev.\ {\bf D50}, 1173 (1994).
\item S.G. Kamath, Nucl.\ Phys.\ {\bf B198}, 61 (1982).
\item J.O. Eeg, Z.\ Phys.\ {\bf C21}, 253 (1984).
\item P. Zenczykowski, Phys.\= Rev.\ {\bf D44}, 1485 (1991) and update quoted in
Ref.~4.
\item G. Eilam, A. Ioannissyan, R.R. Mendel and P. Singer, Phys.\ Rev.\ 
{\bf D53}, (1996).
\item N.G. Deshpande, X.-G. He and J. Trampetic, Phys.\ Lett.\ {\bf B367}, 362 (1996);
see also G. Ricciardi, Phys.\ Lett.\ {\bf B355}, 313 (1995).
\item F. Gilman and M.B. Wise, Phys.\ Rev.\ {\bf D19}, 976 (1979).
\item R. Safadi and P. Singer, Phys.\ Rev.\ {\bf D37}, 697 (1988); 
{\bf D42}, 1856 (E) 1990.  Note that a factor of
$\frac{1}{16}$ is missing in eqs.~(26) and (27).
\item N. Vasanti, Phys.\ Rev.\ {\bf D13}, 1889 (1976).
\item P. Singer, Phys.\ Rev.\ {\bf D42}, 3255 (1990).
\end{enumerate}
\end{document}